# Impacts between multibody systems and deformable structures

Krzysztof Lipinski

*Faculty of Mechanical Engineering and Ship Technology; Gdansk University of Technology; Gdansk; Poland*

klipinsk@pg.edu.pl;  https://orcid.org/0000-0002-7598-4417

**Abstract:** Collisions and impacts are the principal reasons for impulsive motions, which we frequently see in dynamic responses of systems. Precise modelling of impacts is a challenging problem due to the lack of the accurate and commonly accepted constitutive law that governs their mechanics. Rigid-body approach and soft contact methods are discussed in this paper and examined in the presented numerical examples. The main focus is set to impacts in systems with multiple unilateral contacts and collisions with elastic elements of the reference. Parameters of interconnecting unilateral springs are under discussion.

*Keywords: Impacts; Multibody and elastic structures; Contact modelling*

## Statements and Declarations

No funding was received to assist with the preparation of this manuscript. The author has no relevant financial or non-financial interests to disclose.

## 1. Introduction

The final target point of the presented research leads us to bio-inspired mobile robots, especially those able to reconstruct the natural mobility of gibbons. The principal mode of their locomotion is called brachiation. It consists of swinging from branch to branch for distances of up to 15 m and at speeds up to 50 km/h (Fig. 1). We may address the readers to several brachiation techniques and constructions presented in the technical literature [1-5]. Seeing several similarities, we may classify the brachiation robots as a branch of the walking ones (Fig.1a). Each research on the brachiation dynamics is challenging, mainly because of their multitasking: the system's number of degrees of freedom varies during the motion (i.e., we need model a nonlinear time-varying system), unilateral constraints are present (i.e., impact forces can appear) at selected stages of their locomotion, the investigated systems are kinematically or dynamically overactuated. What's more, if a real-world implementation is under investigation, it requests the use of developed intelligent and robust control schemes.

Of course, commercial use of brachiation robots is low and will remain low. We can use them to detect faults in voltage transmission lines. They can inspect and clean electrical cables (e.g., shake off icing). They can work at a high voltage or at heights that are dangerous for humans. What is more, brachiation is not a simple method of locomotion. Its particularly complex aspect is the control. To grasp the cable, the robot must move its temporarily released gripper to the spatial position occupied by the cable to be grasped (Fig. 1b). Since the cable is a vibrating deformable system, the actual position of the grasped point is not a priori known, i.e., it is not a constant of the system. The control system must correctly anticipate the appropriate moment and place of grip activation. But the other "hand" of the robot is in contact with the same vibrating cable. Accordingly, the two subsystems interact bilaterally. Any action of the robot (e.g., a control-inspired one) affects the beam dynamics, and any beam dynamics affects the robot motion, e.g., the impact appearing at the contact point can significantly disorder the motions of both subparts of the system. The challenge is to find an effective method to control this type of robot. Three techniques are predominant: a visual control (it requires a significant number of installed optical detectors), experimental identification of the system behaviour, and numerical modelling of its dynamics. However, if we devote the robot to brachiate in an unexplored environment, experimental identifications are not accessible, and thus, we limit the attention of this article to the numerical modelling of the process. What is more, we are not able to present all the aspects in a single paper. In the present one, we focus on modelling the contact impacts appearing when the robot touches the cable.

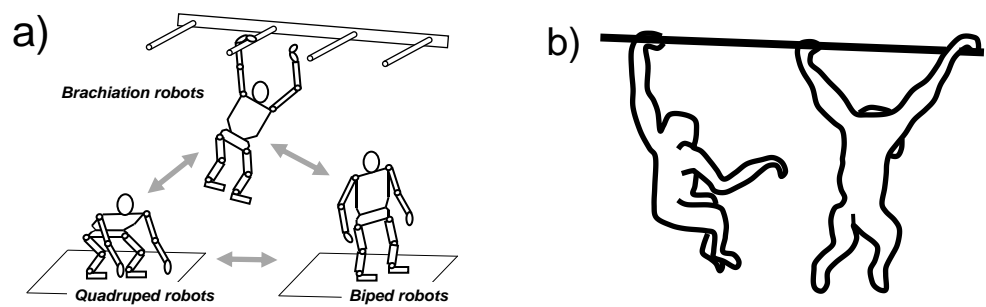

**Fig. 1** Bio-inspired locomoion: three classic types of bio-inspired locomotions (a); selected steps of brachiation (b)

It leads us to a more general problem since collisions and impacts are not only associated with the brachiation. We shall treat them as the principal reason for impulsive motions observed in dynamic responses of mechanical systems. Clearances at mechanical joints, gear train backlashes, motion limiters, capturing of objects, ratcheting mechanisms, and walking or brachiation robots are some examples. Precise modeling of impacts is challenging, mainly due to the lack of the accurate and commonly accepted constitutive law that governs impact mechanics. As pointed out in [6], to model the impact, two alternative attempts are dominant: the impulse/momentum-balance law (a) or constitutive formulas that govern the mechanics of deformations at the small zones of contact and taking place in the short-term interval (b). With the first method, called the *rigid body approach* [6], *impulsive constraints* [7], or *rigid contact* [8], researchers omit details of the evolution of the contact forces. Instead, they balance the velocity changes with the impulses of forces. With the use of the second one, called the *soft contact* [8], researchers investigate details of local deformations. Therefore, we need accurate models of these local deformations, but the lasts are problematic to develop. According to the limited sizes of specimens and rapidity of the processes, experimental measurements are practically impossible and must be replaced by the theory-driven research. Eventual verifications must be limited to some global behaviors only. There are also some ideas of mixed approaches that combine elements applied from both the introduced methods, but they will not be the points of investigation presented in the present paper.

As reported in [9], the above techniques can be applied straightforwardly only if a single unilateral contact is present in the system. If we focus on unilaterally multi-constrained systems, then (except for the principal impulses resulting from the collision) impulse responses may occur at other constraints (but not necessarily at all). Since we do not know the order of the contact vanishing, we cannot apply the momentum balance method directly [9]. The other problem is the need to express details of the energy transfers, e.g., the amount of energy accumulated in the post-impact vibrations remaining in the elastic areas of the contact [10] (generally, it is assumed that the amount of the vibration-converted energy is negligible). The present paper attempts to go beyond these assumptions and present a critical look at the problem of modelling impulsive interactions appearing in multibody systems colliding with elastic structures.

Formally, collisions are forms of contact, and we may model them as constraints. It obligates us to operate with less convenient unilateral constraints and leads us to certain scientific structures classified as *non-smooth dynamics problems* [11-14] and/or *dynamics with discontinuous events* [6]. Most of the investigated cases operate with scleronomous versions of the constraints. Unilateral, rheonomous constraints are seldom [15-17].

Dynamics of impacts is not the principal branch in multibody dynamics, but it has been recognized and investigated from the early beginning. In the 70's of the previous century, Wittenburg [18] used the Newton-Euler dynamics equations to investigate impacting bodies. He operated with a rigid contact model and restricted his studies uniquely to the normal forces. Impulses of the expansion phase were modeled by the use of the restitution coefficient. Also, Moreau [19, 20] and Panagiotopoulos [21, 22] are recognized as pioneers of impact analyses. Haug et al. [23] applied the virtual work principle in investigations focused on impacts within multibody systems. Khulief and Shabana [24], as well as Rismantab-Sany and Shabana [25], operated with the generalized momentum balance method. They have examined the differences between regular and lumped mass formulations in the case of flexible multibody systems. Lankarani and Nikravesh [26, 27] addressed their attention to models of contact forces, especially those resulting from Hertz's contact law. They used canonical impulse-momentum equations for their impact analysis. To model unilateral contacts, Glocker and Pfeiffer [28, 29] referred to the principles of the linear complementarity problems and announced some differences between results when dealing with Newton's and Poisson's impact laws. Chang and Huston [15] used Kane's method to model impacts in unconstrained multibody systems. Zakhariev [30, 31] examined impacts extended with Coulomb friction forces. Ebrahimi and Eberhard [32, 33] focused on frictionless and frictional impacts applied to planar deformable bodies. They formulated the impact problem of continua as a linear complementarity problem on the position level based on the Signorini conditions. Stronge [34] used the energetic coefficient of restitution to operate with frictional collisions. Ambrosio [35] applied a finite-element continuous force model, and appropriate/developed contact models to investigate impacts. Müller [14] focused on time integration of dynamics of variable-topology mechanisms. He combined the momentum balance with the kinematic compatibility conditions to evaluate the velocity changes. Chadaj et al. [36] proposed a formulation based on Hamilton's canonical equations. Accordingly, he operated with the balance of generalized momentum and force impulses at all stages of the performed calculations. Schreyer and Leine [37] presented a mixed shooting–harmonic balance method applied to a large linear mechanical system with local nonlinearities. They modeled the unilateral constraints according to the idea of rigid contact accompanied by the Newton impact law. Dupac [38] studied the dynamics of a spatial impact present on an external surface hit by a rigid beam attached to a sliding structure. He calculated the normal impulsive forces by coupling the elastic-plastic indentation theory with the standard Hertzian contact theory. Jankowski et al. [39] investigated the applicability of their novel contact-force model for selected materials and body shapes and established a strategy to identify parameters appearing in the proposed contact-force expression. Tschigg and Seifried [40] employed a dedicated type of surface finite element to model the contact. To decrease the complication of the proposed model and highlight low-frequency phenomena (fundamental in terms of wave propagation), they damped the high-frequency modes but kept the low-frequency modes as not damped.

A quick overview of the bibliographic positions [41-45] allows us to express a conclusion that if a frictionless point contact connects a multibody and a finite-elements subpart, the most classic and popular modeling method also is the lumped-mass spring or spring/damper connecting element. Accordingly, the numerical integration of the two subparts is quasi-independent, i.e., separate algorithms can calculate their dynamics with the contact force calculated outside. At any request, it calculates the relative penetration. Using a priori assumed elasticity parameters, it estimates contact force and adds it to the dynamics of both the subparts as an external force. The proposed method is attractive. However, from a mathematical point of view, this approach is a version of the classical penalty method. The presented description refers to its canonical version, but modified versions were proposed in the literature, also. In [46], the authors operate with the relative penetrations but use them to evaluate strains that shall be applied in the finite elements of the elastic counterpart. Antunes et al. [47] adopted the Hertzian model of contact force (with some additional damping hysteresis) to evaluate the surface-orthogonal component of the contact force. In [44], a wide list of penalty functions was presented and tested numerically. Afterward, Skrinjar et al. extended the list in [48]. In [49], Atanasovska proposed an extended set of bodies to model the contact. She introduced an additional fictitious deformable body located in a fictitious free zone that separates the contacting physical bodies and treated it as an element that connects them and to them through a set of time-varying dampers and springs.

The present paper is composed of eight sections. Section two presents the fundamentals of the used multibody formalism. Section three focuses on frictionless impacts. A method proposed there balances the impacts of forces and velocity changes. Section four extends it to frictional cases. Section five presents a numerical example of the application of these methods. Section six discusses problems observed in case of impacts present in systems with multiple unilateral constraints. The paper's seventh section discusses the effects of collisions with elastic elements of the reference. Section eight presents conclusions.

## 2. Employed multibody formalism

To develop numerical models, we employed the classical method proposed in [50-52]. We focus on behaviors of strictly *multi-rigid-body systems* (*MBS*). Investigated bodies (Fig. 2a) are inertial and can significantly change their relative positions and orientations. We may lock some of the relative degrees by massless *physical constraints*. According to [51-53], one-degree-of-freedom *joints* (prismatic or revolute type) are sufficient to model such systems. We may also use these joints to introduce propulsion, damping, and elasticity features to the system. We use joint relative displacements (*joint coordinates*) to express poses of the system, i.e., we assume them as the *system coordinates* (*SC*), and we collect them in a single column matrix, **q**.

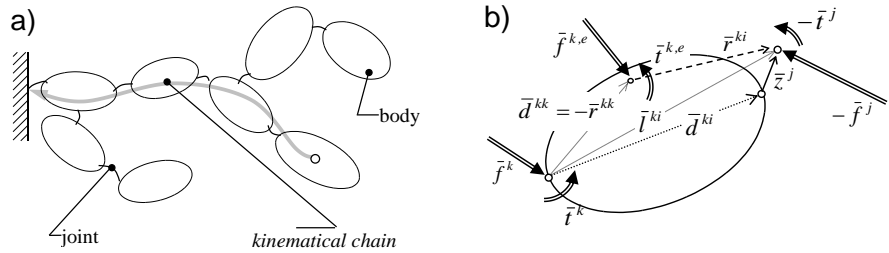

**Fig. 2** Exemplary multibody system: elements of the considered tree-like multibody system (a); geometrical dimensions and interactions present and acting on the cut out body #*k* (b)

If focusing on the methodology, its constitutive structures are the *tree-like multibody systems* (Fig. 2a). We may extend them with *constraint equations* (and their associated *constraint interactions*). With the tree-like structures at hand, their joint coordinates are independent and sufficient to unambiguously describe all system poses. We use orientation matrices, $\boldsymbol{A}^i$, to express the body distances in the user-selected collective reference system. Accordingly:

$$\boldsymbol{A}^i = \prod_{k:k \leq i} \boldsymbol{R}^k \quad ; \quad \bar{x}^i = \sum_{k:k<i}\left(q^k \cdot \bar{a}^k + \bar{d}^{ki}\right) = \sum_{k:k<i} \bar{l}^{ki} \quad (1)$$

where: $\boldsymbol{R}^k$ – relative orientation matrix for bodies interconnected by joint #$k$; $\bar{x}^i$ – absolute position of the mass centre of body #$i$, measured from the origin of the reference system; $\bar{a}^k$ – unit vector expressing direction of the translational motion of translational joint #$k$ (it is a nonzero one for translational joints and zero one for rotational ones); $\bar{d}^{ki}, \bar{l}^{ki}$ – principal relative distances of body #$k$ (see Fig. 2b)

According to the formulae presented in [50-52], we calculate time derivatives of the position vectors and orientation matrices to obtain velocities and accelerations formulae (linear and angular). Following the idea of tables of vectors (*vetrices*) introduced in [50], we write them as [50]:

$$\dot{\bar{x}}^i = \overline{\boldsymbol{A}}^{1,i} \cdot \dot{\boldsymbol{q}} \quad ; \quad \bar{\omega}^i = \overline{\boldsymbol{A}}^{2,i} \cdot \dot{\boldsymbol{q}} \quad ; \quad \ddot{\bar{x}}^i = \overline{\boldsymbol{A}}^{1,i} \cdot \ddot{\boldsymbol{q}} + \ddot{\bar{x}}^{i,R} \quad ; \quad \dot{\bar{\omega}}^i = \overline{\boldsymbol{A}}^{2,i} \cdot \ddot{\boldsymbol{q}} + \dot{\bar{\omega}}^{i,R} \quad , \quad (2)$$

where: $\overline{\boldsymbol{A}}^{1,i}, \overline{\boldsymbol{A}}^{2,i}$ – row vetrices that collect partial velocity vectors for linear and angular velocities respectively; $\ddot{\bar{x}}^{i,R}, \dot{\bar{\omega}}^{i,R}$ – terms of the acceleration vectors uniquely dependent on the velocity products.

When deriving the *dynamics equations*, all joints are cut out and replaced by the joint interactions (forces and torques), i.e., we draw free-body diagrams for all bodies of the structure (see Fig. 2b). Then, we use Newton/Euler dynamics equations to describe their dynamics, and we combine the results with the vetrix-based formulae (2). To eliminate the direct-successor's forces and torques from the equations, we use the backward evaluation technique. After a rearrangement, we may write the formulae for interactions at the cut joint #$i$ as [50]:

$$\bar{f}^i = \overline{\boldsymbol{C}}^{1,i} \cdot \ddot{\boldsymbol{q}} + \bar{d}^{1,i} + \bar{e}^{1,i} \quad ; \quad \bar{t}^i = \overline{\boldsymbol{C}}^{2,i} \cdot \ddot{\boldsymbol{q}} + \bar{d}^{2,i} + \bar{e}^{2,i} \quad , \quad (3)$$

where: $\overline{\boldsymbol{C}}^{1,i}, \overline{\boldsymbol{C}}^{2,i}$ – row vetices that collect partial force and torque vectors respectively; $\bar{d}^{1,i}, \bar{d}^{2,i}$ – terms dependent on the velocity products, developed for force and torque, $\bar{e}^{1,i}, \bar{e}^{2,i}$ – net effects calculated from the external forces and torques acting at all the bodies located above of the cut joint #$i$.

Vectors (3) are projected on the joint mobility vectors (for translational joints, the cut-joint force, $\bar{f}^i$, is projected on, $\bar{a}^i$, vector, and for the revolute ones, the cut-joint torque, $\bar{t}^i$, is projected on, $\bar{e}^i$, vector). Finally, components in front of the joint accelerations are collected in the mass matrix. The result is [50-52],

$$M(q) \cdot \ddot{q} + F(\dot{q}, q, f_e, t_e, t) = Q \quad , \qquad (4)$$

where: $M$ – mass matrix ($n \times n$ matrix); $F$ – column matrix ($n \times 1$ matrix) composed of velocity-product inertial terms and of external forces and torques; $Q$ – column matrix composed of joint actuations (propulsions); $f_e$ – column matrix composed of the external forces (acting at the bodies of the system); $t_e$ – column matrix composed of the external torques (acting at the bodies of the system); $t$ – time. The introduced algorithms/formulae are general, and we may use them for the planar or 3D motions of the elements. To minimize the probability of a potential error, we use a computer algorithm to obtain components of the matrices from (4). The algorithm works according to a methodology presented in [51, 52].

## 3. Impacts in frictionless multibody systems

Let us look at a frictionless collision. An element of a multibody system impacts the reference body. In the initial state of motion, its configuration is free of contact. During impact, a contact force, $f_c$, is added at point, $C$, of the multibody structure. To express the presence of the force, we shall modify (4) to,

$$M(q,t) \cdot \ddot{q} + F(\dot{q}, q, f_e, t_e, t) + J_c^T \cdot f_c = Q \qquad (5)$$

where: $J_c^T$ – Jacobian of the formula that express the absolute position of C. Since we investigate collisions of rigid elements, the surface-normal component of the approaching velocity reduces to zero during the infinitesimal time of the impact (any nonzero velocity will result in a penetration inside of the colliding bodies). Finite forces are ineffective since infinite collision forces can only create infinite accelerations. Such events are not predisposed to numerical integrations, i.e., we shall stop the integrations a moment before the collision. Even though the accelerations are infinite, their integrals (i.e., velocity changes) are finite. Accordingly, to obtain the joint's velocities changes, we shall study the integrated form of the dynamic eq. (5), (i.e., the balances of impulses and the velocity changes [6, 9, 14, 50], or impulses and momentums [36]). It leads to,

$$\int_t^{t+\Delta t} M(q,t) \cdot \ddot{q} \cdot dt + \int_t^{t+\Delta t} F(\dot{q}, q, f_e, t_e, t) \cdot dt + \int_t^{t+\Delta t} J_c^T \cdot f_c \cdot dt = \int_t^{t+\Delta t} Q \cdot dt \qquad (5)$$

where $\Delta t$ is the infinitesimal period of the impact. According to their infinitesimal durations, integrals of finite quantities (e.g., $\boldsymbol{F}$ and $\boldsymbol{Q}$) are negligible. Only integrals of the accelerations and contact force must be under the studies. Moreover, remembering that position changes are integrals of finite velocities, all the joint positions are constant, i.e., all the matrices of (5) should be assumed constant. Applying these assumptions, we can simplify (5) to [6, 9, 14, 50],

$$\boldsymbol{M}(\boldsymbol{q},t) \cdot \Delta \boldsymbol{q} + \boldsymbol{J}_e^T(\boldsymbol{q}) \cdot \boldsymbol{I} = 0 \quad ; \quad \Delta \boldsymbol{q} = -\boldsymbol{M}^{-1} \cdot \boldsymbol{J}_e^T \cdot \boldsymbol{I} \quad , \tag{6}$$

where:
$$\Delta \boldsymbol{q} = \int_t^{t+\Delta t} \ddot{\boldsymbol{q}} \cdot dt \quad ; \quad \boldsymbol{I} = \int_t^{t+\Delta t} \boldsymbol{\lambda}_e \cdot dt \quad . \tag{7}$$

The joint-velocity changes must comply with the kinematics of the systems. The contact begins with a non-zero surface-normal component of the approaching velocity, i.e., the joint velocities, $\dot{\boldsymbol{q}}^b$, and the velocity of the multibody point, $\boldsymbol{v}_b$, are non-zero. At the end of the compression, the joint velocities evolve to altered non-zero values, $\dot{\boldsymbol{q}}^c$, but the motion of the multibody point stops. Thus,

$$\boldsymbol{J}_e \cdot \Delta \dot{\boldsymbol{q}} = \boldsymbol{J}_e \cdot \left( \dot{\boldsymbol{q}}^c - \dot{\boldsymbol{q}}^b \right) = \boldsymbol{J}_e \cdot \dot{\boldsymbol{q}}^c - \boldsymbol{J}_e \cdot \dot{\boldsymbol{q}}^b = 0 - \boldsymbol{J}_e \cdot \dot{\boldsymbol{q}}^b . \tag{8}$$

If changes of the joint velocities are calculated from (6b),

$$-\boldsymbol{J}_e \cdot \left( \boldsymbol{M}^{-1} \cdot \boldsymbol{J}_e^T \cdot \boldsymbol{I}_c \right) = -\boldsymbol{J}_e \cdot \dot{\boldsymbol{q}}^b , \tag{9}$$

and the impulses of the compression phase can be calculated as [9, 50],

$$\boldsymbol{I}_c = \left( \boldsymbol{J}_e \cdot \boldsymbol{M}^{-1} \cdot \boldsymbol{J}_e^T \right)^{-1} \cdot \boldsymbol{J}_e \cdot \dot{\boldsymbol{q}}^b . \tag{10}$$

Assuming the existence of dissipate properties of the contacting substances, impulses of the expansion phase are modelled as fractions of impulses (10) (impact dissipation), expressed by a restitution coefficient, $R$. It leads to [9, 50],

$$\boldsymbol{I}_e = R \cdot \left( \boldsymbol{J}_e \cdot \boldsymbol{M}^{-1} \cdot \boldsymbol{J}_e^T \right)^{-1} \cdot \boldsymbol{J}_e \cdot \dot{\boldsymbol{q}}^b , \tag{11}$$

and the total changes of the joint velocities are [9, 50],

$$\Delta \dot{\boldsymbol{q}}_\Sigma = -\boldsymbol{M}^{-1} \cdot \boldsymbol{J}_e^T \cdot \left( \boldsymbol{I}_c + \boldsymbol{I}_e \right). \tag{12}$$

## 4. Impact tangent components

The presence of friction indicates additional complexities and connections. Since we calculate values of developed friction forces using the coefficient, $\mu$, thus the tangent and normal impulses are also correlated. Moreover, when the normal impulse changes the joint velocities, it modifies both the velocities: the surface-normal and tangent components of the multibody-point velocity. Similarly, the tangent impulse also modifies both components. In the planar case, it leads to:

$$\Delta v_n^f = C_{11} \cdot I_n + C_{12} \cdot I_t \; ; \quad \Delta v_t^f = C_{21} \cdot I_n + C_{22} \cdot I_t \; ; \quad \boldsymbol{C} = \boldsymbol{J} \cdot \boldsymbol{M}^{-1} \cdot \boldsymbol{J}^{\mathrm{T}}, \quad (13)$$

where: $\Delta v_n^f$, $\Delta v_t^f$ – change of the normal and the tangent component of the velocity; $I_n$, $I_t$ - impulse of the normal and the tangent force, respectively.

Eqs. (13) contain four unknown: $I_n, I_t, \Delta v_n^f$, and $\Delta v_t^f$. Additional equations/conditions are necessary to solve them. To step forward, we need to identify the case dedicated order of two critical time instants, i.e., $t_n$ – end time of the compression phase; $t_t$ – end time of the slip phase.

Firstly, we shall test if the friction can stop the slip. When assuming a permanent slip, the tangent impulse is a fraction of the normal impulse (we have supposed that the classical model of Culomb is applicable for impact forces. It is disputable). Accordingly, we can rewrite Eq. (13) as:

$$\Delta v_n^{st} = (C_{11} + s \cdot \mu \cdot C_{12}) \cdot I_n^{st} \; ; \quad -v_t^b = (C_{21} + s \cdot \mu \cdot C_{22}) \cdot I_n^{st}, \quad (14)$$

where: $I_n^{st}$ – impulse of the normal force needed to stop the slip; $s = \mathrm{sign}(v_t^b)$ – initial direction of the tangent slip; $v_t^b$ – initial value of the slip velocity; $\Delta v_n^{st}$ – change of the velocity normal component.

Eqs. (14) contain two unknown: $I_n^{st}$ and $\Delta v_n^{st}$, they shall be solved and tested. Firstly, if $I_n^{st}$ is negative, the slip accelerates during the impact (i.e., friction is insufficient to reduce the slip). Eqs. (13) are useless, we shall replace them by eqs. (14). Secondly, if $I_n^{st}$ is positive, we shall next focus on the next set of equations:

$$-v_n^b = (C_{11} + s \cdot \mu \cdot C_{12}) \cdot I_n^c \; ; \quad \Delta v_t^c = (C_{21} + s \cdot \mu \cdot C_{22}) \cdot I_n^c, \quad (15)$$

It allows us to evaluate the total normal impulse of the compression, $I_n^c$, and the change of the slip velocity, $\Delta v_t^c$. The slip velocity at the end of this phase is

$$v_t^c = v_t^b + \Delta v_t^c, \quad (16)$$

The slip velocity (16) is lower than the initial (since the $I_n^{st}$ is not negative). If the velocity changes its sign, it indicates the stop of the slip during the compression, i.e., the inequality, $I_n^{st} < I_n^c$, is true, and eqs. (15), (16) are useless. The first instant is the stop of the slip, and we shall evaluate its impulses from (14). Again, two kinds of future velocity evolutions are possible. Firstly, the future slip can be locked. The residual impulses, $I_n^{rc}$ and $I_t^{rc}$, of the compression shall be calculated from:

$$-(v_n^{st} + \Delta v_n^{st}) = C_{11} \cdot I_n^{rc} + C_{12} \cdot I_t^{rc} ; \qquad 0 = C_{21} \cdot I_n^{rc} + C_{22} \cdot I_t^{rc}. \quad (17)$$

Next, we shall compare the tangent impulse, $I_t^{rc}$, with its limit value, $\mu \cdot I_n^{rc}$. If the calculated one is higher, we may not preserve the assumption about the stop of the slip. We have to reject eqs. (17), and replace them by formulae:

$$-(v_n^{st} + \Delta v_n^{st}) = (C_{11} - s \cdot \mu \cdot C_{12}) \cdot I_n^{rc} ; \quad \Delta v_t^c = (C_{21} - s \cdot \mu \cdot C_{22}) \cdot I_n^{rc}, \quad (18)$$

Since the initial configuration of impulses reduces the tangent speed, they may not accelerate it now. It indicates the inverse of the slip and the tangent impulse. Similar equations shall be tested for the expansion phase (assuming that the final value on the normal impulse is expressed by the restitution coefficient).

## 5. Planar system with rigid impact and friction

The tested manipulator consists of 10 bodies connected by revolute joints. Its endpoint collides with a rigid reference/obstacle. We model the obstacle as a stationary vertical line. After the impact, the manipulator bounces off. All segments of the manipulator have identical dimensions and inertial parameters. Their lengths are 0.17 m, masses are 0.3341 kg, and moments of inertia respectively to the centre of mass are 0.0011388 kg·m². All drives at the joints are unlocked. Gravity forces are excluded. In its initial pose, displacements in all joints equal -0.25 rad, and the initial velocities are -0.2 rad/s. Slides displaying selected phases (before and after impact) are shown in Fig. 3. The figure presents results obtained with different friction coefficients occurring in the contact area.

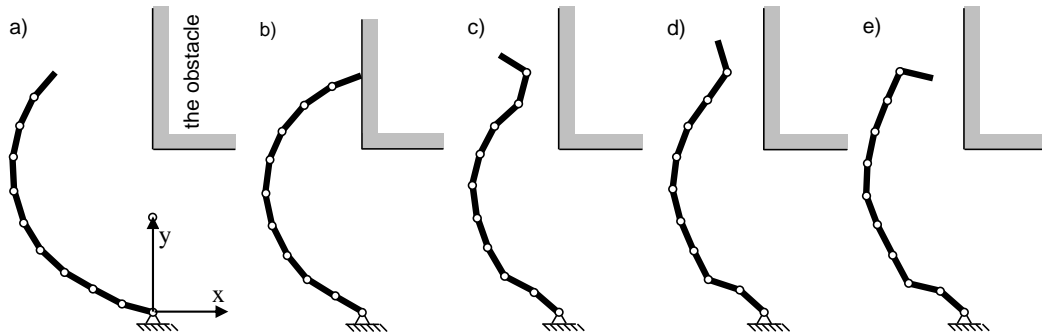

**Fig. 3** Selected slides showing the motion of the manipulator: pre-impact pose, $t = 0.0$ s (a); impact pose, $t = 0.390$ s (b); post-impact pose, $\mu = 0.0$, $t = 0.561$ s (c); post-impact pose, $\mu = 0.15$, $t = 0.561$ s (d); post-impact pose, $\mu = 0.35$, $t = 0.561$ s (e)

# 6. Planar system with multiple friction-free impacts

We composed the tested system of two material bodies (see Fig. 4ac): the main body #3 and the moving arm #4. The main body is connected with the reference using a sequence of three joints (linked by two point-like massless bodies #1 and #2). The used joint sequence is T1/T2/R3 (horizontal/vertical/rotation). The low corners of the main body are fixed to the reference by unilateral contact constraints (points *A* and *B* in Fig. 4a and constraints $sc_A$ and $sc_B$ in Fig. 4c). The mass of the main body is $m_3 = 10$ kg. Its moment of inertia equals $I_3 = 0.1$ kg·m². The position of the mass center coincides with the revolute joint connecting the main body with the reference. The main geometrical parameters of the body and the contact points are presented in Fig. 4b. The moving arm #4 is connected to the main body using a revolute joint. It rotates at a relatively high speed. Inertial terms (e.g., the d'Alembert forces) are not high (in comparison to the gravity of the main body) and insufficient to interrupt contacts at points *A* and *B*. We fix the arm to the main body at the local coordinates that equal $l_{34x} = 0.15$ m and $l_{34y} = 0.1$ m (with respect to the mass center of the main body). Its mass equals $m_4 = 2$ kg, and the moment of inertia is $I_3 = 0.043$ kg·m². The position of its mass center is $l_{44x} = 0.2$ m and $l_{44y} = -0.08$ m (respectively to the joint connecting the arm with the main body). The coordinates of the impact point, *C*, are $l_{4Cx} = 0.3$ m and $l_{4Cy} = -0.2$ m (respectively to the joint connecting the arm with the main body).

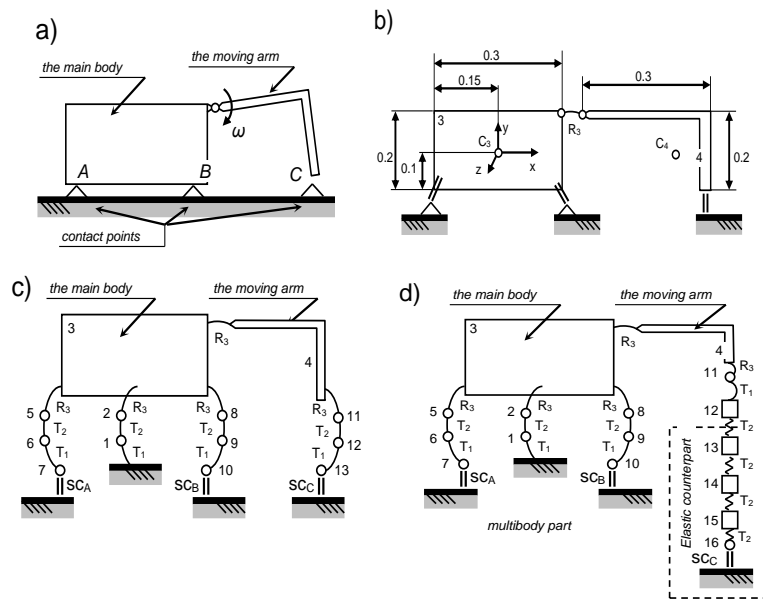

**Fig. 4** Structure of the investigated system: reference structure (a); main geometrical parameters (b); multibody structure – arm impacting on motionless reference (c); multibody structure – arm impacting on deformable structure (d)

Let us investigate the impact of the moving arm and the ground. When point $C$ at the end of the moving arm reaches the ground (its vertical coordinate $y_g$ is assumed as equal zero), the two main body points also contact the ground (i.e., points $A$ and $B$ in Fig. 4a). We use the smoothed contact model (elastic contact area). Obtained dynamics equations are investigated numerically in MATLAB [53]. In the system's initial pose, we put the main body horizontally. Initial rotation angle moving arm is $q_4(0) = 2$ rad, and its rotational speed is $qp_4(0) = -20$ rad/s. We used the ODE45 procedure for numerical integration, and limited the maximal and initial time steps to $10^{-4}$ s. The relative accuracy is set to $10^{-4}$, and the absolute one to $10^{-6}$.

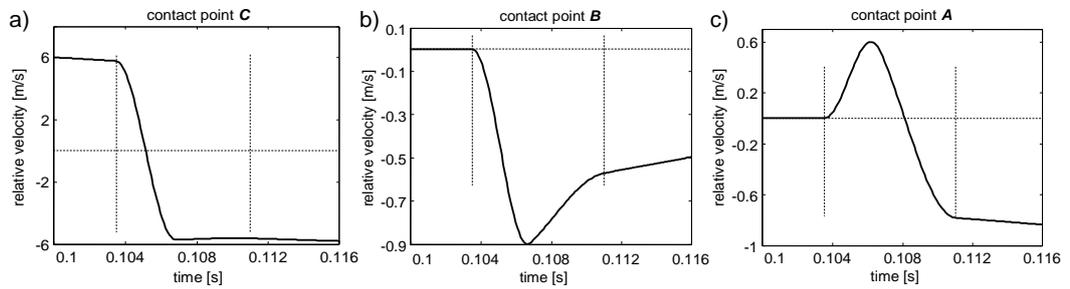

**Fig. 5** Velocities of the relative penetrations: point $C$ (a); point $B$ (b); point $A$ (c)

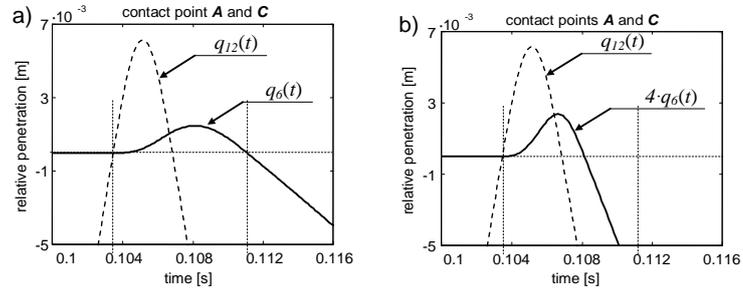

**Fig. 6** Relative penetration: $c_{1A}= 10^6$ N/m (a); $c_{2A}= 4 \cdot 10^6$ N/m (b)

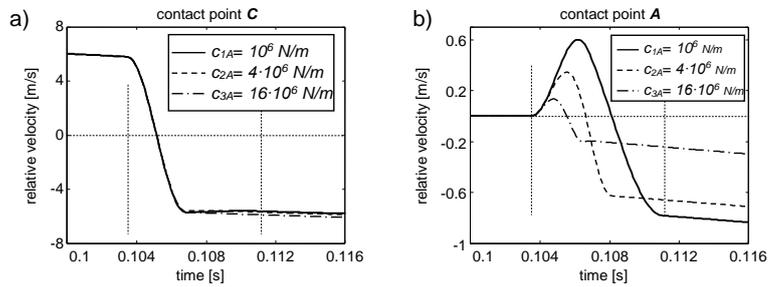

**Fig. 7** Velocities of the relative penetration: point $C$ (a); point $A$ (b);

Initially, identical elasticity coefficients of $10^6$ N/m are supposed for the contact points, and the relative penetrations are investigated for the three contact points. As we can see in Fig 5a, the co-impact only takes place in point $A$. Point $B$ is free of the co-impact (Fig.5b). From the beginning, its velocity evolves to negative values (Fig. 5b). The main impact at point $C$ results in about 3 times higher penetration than in the co-impacting point *A*, but its duration is about 2 times shorter than the duration at point *A* (Fig. 6a).

In the subsequent tests, we tested different values of the elasticity coefficients applied at joint #6 (at point *A*): $c_{1A}= 10^6$ *N/m*, $c_{2A}= 4·10^6$ *N/m*, $c_{3A}= 16·10^6$ *N/m*, (leaving the previous elasticity coefficient of $10^6$ *N/m* at joint #12). We observed relative penetrations (Fig. 6) as well as their velocities (Fig. 7). Displacements at point *C* ($q_{12}$) are almost identical (Fig. 6), i.e., slightly dependent on the elasticity at point *A*. Contrarily, the magnitudes and duration of penetrations at point *A* depend significantly on elasticity. Also, areas surrounded by the positive parts of these characteristics (correlated to impulses of the contact forces) are different. As seen in Fig. 7b, post-impact velocities at point *A* strongly depend on the impacted elasticity. By contrast, post-impact velocities are almost identical at point *C* (see Fig. 7a). We are obligated to underline that the application of the impulse-based analysis (eqs. (5)-(18)) can be problematic in the case of multiple unilateral contacts. The presented results certify that the ratio correlating the elasticity coefficients at the distant contact points is a significant model parameter.

## 7. Impacts with deformable structures

Again, three unilateral contacts (Fig.4a) are considered. Contrary to the previous case, an elastic counterpart is built of 3 masses serially connected by springs (Fig. 4d). Stiffness of the springs is $1.6·10^5$, except for the one between body #12 and #13, which is a unilateral spring, i.e., it can only transfer the compressive forces. Fig. 8 presents displacements/deformations of the investigated colliding unilateral spring for different values of its elasticity. The reference value of its elasticity equals the one used to connect the serial masses (the dotted line in all the subfigures), i.e., $c_{13}= c_{\text{ref}} =1.6·10^5$. All springs are free of damping/dissipation of the energy (we intend to investigate a purely elastic impact). If relatively low elasticity coefficients are assumed (Fig. 8a), we can observe a continuous contact during the whole contact period of 0.02 s. If higher coefficients are investigated, there is a sequence of consecutive contact (rebounds and re-contacts), up to 9 rebounds during the contact period. The total duration of the in-touch period is similar in all these cases, i.e., it slightly depends on the elasticity of the unilateral spring.

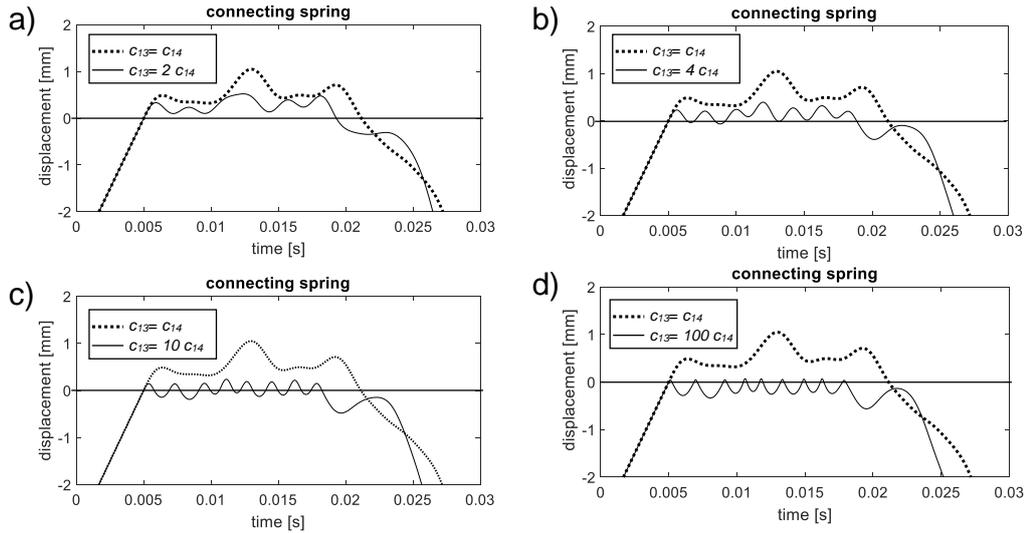

**Fig. 8** Deformations of the investigated unilateral spring: $c_{13} = 2 \cdot c_{ref}$ (a); $c_{13} = 4 \cdot c_{ref}$ (b); $c_{13} = 10 \cdot c_{ref}$ (c); $c_{13} = 100 \cdot c_{ref}$ (d);

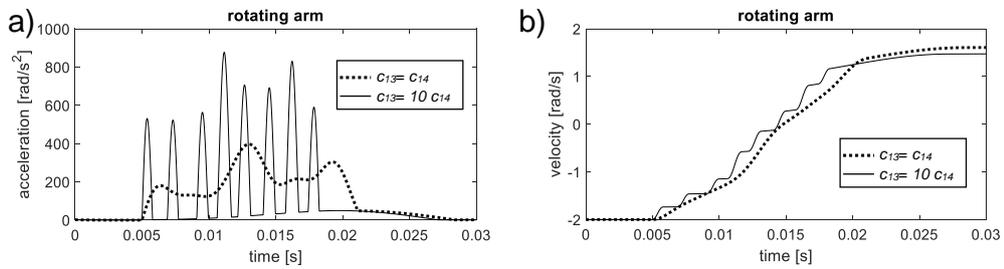

**Fig. 9** Relative kinematics of the rotating arm: accelerations at joint #4; velocities at joint #4;

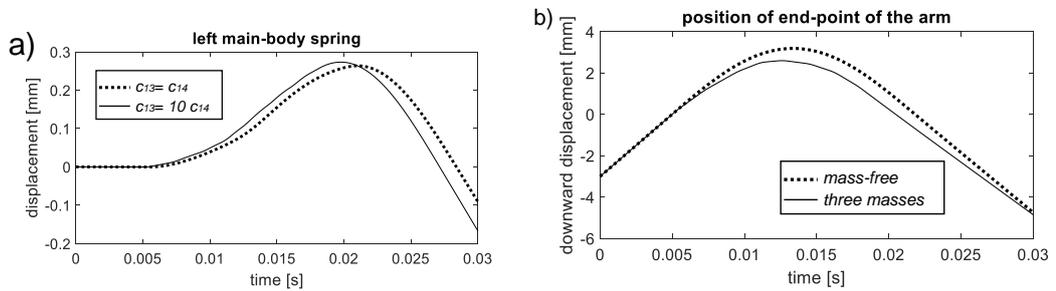

**Fig. 10** Selected displacements: deformation of the left-side unilateral spring attaching the main body to the reference (a); total downward displacement of the arm endpoint impacting on the inertial three-mass structure versus its impact on an inertia-free spring (b);

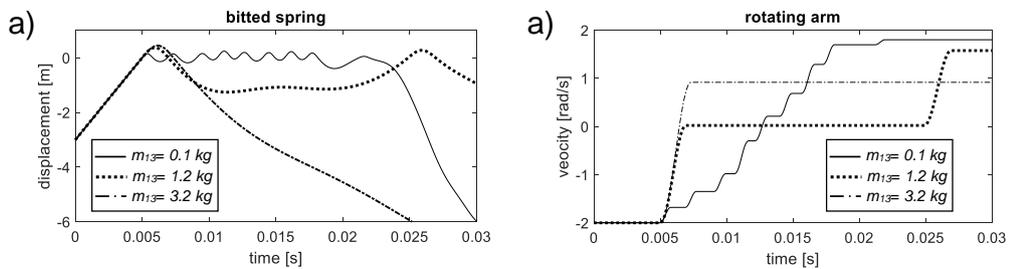

**Fig. 11** Influence of masses of the elastic counterpart: deformation of the left-side unilateral spring fixing the main body to the reference (a); total downward displacement of the arm endpoint impacting on the inertial three-mass structure versus its impact on an inertia-free spring (b);

Fig. 9 presents the joint kinematics of the colliding arm. Two elasticities of the colliding spring are tested: the reference one and the 10 times higher. We can observe the successive rebounds in the accelerations (Fig. 9a). They are also visible in evolutions of the velocities (Fig. 9b), as the rapid changes of the rotational speed. Let us also underline that the observed behavior differs from that typical for elastic rebounds. The final angular velocity is lower than the initial one, i.e., the impacted object does not rebound the total energy back. Its important part remains there as the energy stored in vibrations of the impacted masses.

Fig 10a focuses on the left-hand-side unilateral spring that attaches the main body to the reference (the compressed one). The subsequent rebounds observed at the impacted spring are challenging to detect in Fig. 10a. According to the inertial nature of the main body, the velocity changes resulting from the subsequent impacts are not sufficiently big to be observed in this figure, i.e., presented evolution looks neutral or weakly correlated with values of the impacted elasticity $c_{13}$. Fig. 10b presents the total downward displacement of the arm endpoint impacting on the three-mass structure (i.e., a sum of all the successive joint displacements starting from body #12) and compares it with a resulting of an impact done on an inertia-free single spring of equivalent elasticity. As we can see, both characteristics are comparable, i.e., the influence of masses of the impacted structure is low.

Fig. 11 presents results calculated for different inertia of elements constituting the three-mass system. With the increase of masses of the elastic subpart, we observe three modes of behavior of such systems. The *backward impact* is typical for big masses. The arm impacts once and moves backward at a significant retrograde speed. The *backward+ impact* is typical for intermediate masses. After the first impact, the arm acquires a slight backward velocity, but it does not prevent the subsequent collision of the arm with the faster upper mass of the vibrating flexible counterpart. Finally, if one investigates small masses, a *forward impact* is observed. The arm rebounds the upper mass of the elastic counterparts. It reduces the speed of the arm, but not enough to develop the backward motion. The return of the vibrating mass results in a repeated impact, but again, insufficient to reverse the direction of the arm's speed. A long sequence of collisions is necessary to stop and reverse the direction of the rotation of the biting arm. Accordingly, we can observe a long sequence of multiple rebounds.

## 8. Conclusions

Collisions in multibody systems are not straightforward for modelling, especially in multi-contact cases and in cases of impacts with elastic structures. These last have tendencies to vibrate. Thus, multi-rebound behaviours are typical. The classic unilateral constraint models and impact-momentum balance techniques are impractical in these cases. Since each of the successive impacts demands a stop of the numerical integration, and then in algebraic estimation of the post-impact velocities, the presence of the elastic counterpart enforces us to stop the integration not a single time but several times at each rebound observed in the model. It makes the algorithm complex and less effective. The application of the "soft contact" model is also problematic. The total duration of the in-touch period slightly depends on the value of the connecting unilateral elasticity. Its duration corresponds mainly with the slowest frequency of vibrations of the elastic counterpart. Enlargements of the connecting-spring elasticities result primarily in increases in the number of observed subsequent rebounds. The overall motion of the multibody part appears to be not influenced (or slightly influenced) by assumed values of the connecting elasticity.